\input{aipcheck}

\documentclass[
  ]
  {aipproc}

\layoutstyle{8x11single}
\usepackage{enumerate}
\usepackage{siunitx}
\usepackage{amssymb,amsmath}
\begin{document}

\title{AN EXPERIMENTAL STUDY OF WAVEGUIDE COUPLED MICROWAVE HEATING WITH CONVENTIONAL MUlTICUSP NEGATIVE ION SOURCE}


\author{J. Komppula}{
  address={University of Jyvaskyla, Department of Physics, P.O. Box 35, FI-40014 University of Jyvaskyla, Finland}
}

\author{T. Kalvas}{
  address={University of Jyvaskyla, Department of Physics, P.O. Box 35, FI-40014 University of Jyvaskyla, Finland}
}

\author{H. Koivisto}{
  address={University of Jyvaskyla, Department of Physics, P.O. Box 35, FI-40014 University of Jyvaskyla, Finland}
}

\author{J. Laulainen}{
  address={University of Jyvaskyla, Department of Physics, P.O. Box 35, FI-40014 University of Jyvaskyla, Finland}
}

\author{O. Tarvainen}{
  address={University of Jyvaskyla, Department of Physics, P.O. Box 35, FI-40014 University of Jyvaskyla, Finland}
}

\begin{abstract}Negative ion production with conventional multicusp plasma chambers utilizing 2.45 GHz microwave heating is demonstrated. The experimental results were obtained with the multicusp plasma chambers and extraction systems of the RF-driven RADIS ion source and the filament driven arc discharge ion source LIISA. A waveguide microwave coupling system, which is almost similar to the one used with the SILHI ion source, was used. The results demonstrate that at least one third of negative ion beam obtained with inductive RF-coupling (RADIS) or arc discharge (LIISA) can be achieved with 1 kW of 2.45 GHz microwave power in CW mode without any modification of the plasma chamber. The co-extracted electron to H$^-$ ratio and the optimum pressure range were observed to be similar for both heating methods. The behaviour of the plasma implies that the energy transfer from the microwaves to the plasma electrons is mainly an off-resonance process.
\end{abstract}

\maketitle

%%%%%%%%%%%%%%%%%%%%%%%%%%%%%%%%%%%%%%%%%%%%
%% MAINMATTER
%%%%%%%%%%%%%%%%%%%%%%%%%%%%%%%%%%%%%%%%%%%%

\section{Introduction}
Tandem multicusp ion sources based on volume production through dissociative electron attachment are the most commonly used negative hydrogen ion sources for accelerators \cite{Bacal_volume_production}. The source consists of two plasma regions separated by a magnetic filter field. The plasma is produced and sustained by the main discharge which is typically confined magnetically by a multicusp field. The magnetic filter reduces the plasma temperature due to diffusion in transverse direction across the field. Such cold plasma is suitable for the volume production of negative ions in $e+$H$_2\rightarrow $H$_2^-\rightarrow $H$+ $H$^-$ reaction. Therefore, negative ions are extracted with reasonably low amount of co-extracted electrons. A filament driven arc discharge or inductively coupled RF-discharge driven by an internal or external antenna are typically used to sustain the main discharge \cite{NegativeReview}.

Electron cyclotron resonance (ECR) heating is widely applied for the production of singly and multiply charged positive ions. Both, the most powerful operational proton sources \cite{SILHI} and sources for highly charged ions  \cite{VENUS} utilize ECR heating. Microwaves can also be considered as a reasonable plasma heating method for negative ion sources, potentially providing a maintenance free alternative for filament arc discharges. Furthermore, waveguide coupling of microwaves is less complicated and more reliable than impedance matching circuits required for RF-discharges and the microwave radiation is fully shielded by the waveguide which prevents disturbing the surrounding electronics with high frequency noise. However, to-date microwave heating is not successfully applied to operational negative ion sources.

Several attempts have been made during the last decades to apply microwave heating for negative hydrogen ion production. The earliest study was performed by Hellblom and Jacquot \cite{negative_microwave_first} using 11 GHz electron cyclotron resonance (ECR) heating in solenoid magnetic field. Since then several studies have been performed with different frequencies and types of microwave coupling systems, for example Refs. \cite{ECR_negative_1,ECR_negative_2,ECR_negative_3,ECR_negative_4,negative_microwave_india} and references therein. The main technical challenges in applying microwave heating are related to the magnetic filter field. Conventional magnetic filter field is a dipole field. Superimposing the dipole field with the solenoid field is challenging as it causes significant plasma drifts. To overcome the issue, electrostatic filters have been studied \cite{ECRIN_multicusp_electric_filter} for example.

2.45 GHz plasma sources can be operated with off-resonance magnetic fields (B < 875 mT) as well (e.g. Refs. \cite{non_resonance1,non_resonance2}). In fact, it is possible to operate microwave ion sources without any magnetic confinement but reliable plasma ignition requires resonance conditions in the plasma chamber, which is important for devices operating in pulsed mode. Recently Sahu et al. \cite{negative_microwave_india} demonstrated, that it is possible to produce negative hydrogen ions ($\sim$\SI[per-mode=fraction]{33}{\micro\ampere}, \SI[per-mode=fraction]{0.26}{\milli\ampere\per\centi\meter^2}) with a 2.45 GHz microwave discharge and multicusp plasma chamber, in which the resonance condition is reached close to the plasma chamber wall.

In this study the potential of microwave heating in negative ion production is demonstrated with two different operational CW $H^{-}$ ion sources, RF-driven RADIS ion source \cite{NIBS2012_Kalvas} and filament driven arc discharge LIISA ion source \cite{LIISA}. The plasma heating methods of the given ion sources were transformed to 2.45 GHz microwave heating by replacing the back plates of the plasma chambers with the back plate of a microwave discharge ion source. This provides more reliable comparison of heating methods than a comparison of totally different ion sources, since only minor mechanical changes and variation of tuning are necessary. Because the plasma chambers are not optimized for microwave heating, the presented feasibility study leaves ample
potential for optimization as discussed in the last section of the paper.

\section{Microwave heating}
Microwave-plasma interaction has been studied intensively with microwave discharges. Indications of several different types of wave-particle interactions have been observed, including electron cyclotron resonance (ECR), second order electron cyclotron resonance \cite{non_resonance1,non_resonance2}, upper hybrid resonance \cite{negative_microwave_india} and plasma heating by Bernstein waves \cite{italy_bernstein}. Understanding the microwave-plasma interaction in multicusp plasma chambers is challenging, because the structure of the resonance surface is complex, the plasma is inhomogeneous, the plasma density affects the wavelength of the microwaves and the dimensions of the plasma chambers are typically on the same order of magnitude with the wavelength of the 2.45 GHz microwaves in vacuum. Theories of wave propagation in magnetoplasmas typically assume the plasma dimensions to be large in comparison to the wavelength and, therefore, they should be applied to 2.45 GHz microwave discharges only with extreme caution. The plasma heating by microwaves in absence of magnetic field cannot be described with such theories but should be treated as inductive or capacitive heating (see the discussion for radio frequencies in Ref. \cite{RF_heating_revisited} for example). Therefore in the context of this study, the plasma heating is divided to two categories, which can be observed experimentally in the microwave discharge at issue (Fig. \ref{fig:microwave_performance}):
\begin{enumerate}[i)]
\item\label{item:ECR_heating} Plasma heating in electron cyclotron resonance, which is well-understood.
\item\label{item:nonresonance_heating} Off-resonance heating, which is experimentally observed but only poorly understood.
\end{enumerate}

The efficiency of ECR heating depends on the magnetic field topology and electric field distribution of the plasma chamber. The velocity of an electron bound to magnetic field can be divided to two components, perpendicular $v_{B\bot}$ and parallel $v_{B\parallel}$, with respect to the direction of the field. Electron cyclotron resonance is achieved in the magnetic field, where the electron gyrofrequency is equal to the frequency of the microwaves (B$_{ECR}= 87.5$ mT for  2.45 GHz microwaves). In resonance the perpendicular velocity of the electron $v_{B\bot}$ is increased by the perpendicular component of the microwave electric field  $E_{B\bot}$. 

The magnetic field varies spatially and, therefore, the electron spends only infinitesimally short time in perfect resonance conditions. A first order approximation is that the heating efficiency in off-resonance magnetic field $B$ is proportional to $1/(B-B_{ECR})^2$, which is a simplified form of the general description of the electron heating rate 
\begin{equation}
H\propto\frac{\nu}{(\omega-\omega_e)^2+\nu^2},
\end{equation}
where $\nu$ is the electron collision frequency, $\omega$ the microwave (angular) frequency and $\omega_{e}$ the electron gyrofrequency (see e.g. Ref. \cite{ECR_absorption}). The magnetic field topology of the plasma chamber affects the heating efficiency in two ways. The volume of efficient power absorption can be optimized by maximizing the $1/(B-B_{ECR})^2$ weighted volume of the plasma chamber. On the other hand, the local heating efficiency can be enhanced by maximizing the time spent by the electron near the resonance field. In practice this means minimizing the magnetic field gradient in the direction of the magnetic field lines.

Also the electric field affects the electron heating efficiency. The heating efficiency can be improved by maximizing the electric field at the resonance. In absence of the plasma the plasma chamber acts like a resonance cavity. If the chamber diameter is on the order of $\lambda /2$ where $\lambda$ is the wavelength of the microwave radiation in vacuum, it is probable that the lowest cavity modes are excited also in the presence of plasma. However, if the plasma chamber diameter is significantly longer than the wavelength of the microwaves, plasma affects the cavity modes and, hence, the electric field distribution significantly \cite{VillenKammiot}. Under such conditions it is not reasonable to talk about resonance cavity modes. 

The volume production of negative hydrogen ions requires high densities of low energy electrons and molecules on high vibrational levels, which are produced in the main discharge. Both, ionization and molecule excitation to high vibrational levels via $B^1\Sigma^+_u$ and $C^1\Pi_u$ electronic states, benefit from high electron energies. They are dominant electron impact processes at electron energies exceeding 30 eV \cite{crosssections1}. At lower electron energies, a larger fraction of electron energy is transferred to (VUV) light emission via electron impact excitations from which the most significant are excitations to triplet states \cite{crosssections1}. Those excitations do not contribute to the volume production of H$^{-}$ since the triplet state excitations lead to molecule dissociation via the repulsive $b^3\Sigma^+_u$ state. Furthermore, the electron momentum is changed significantly in electron-electron collisions at low electron energies ($E_e \lesssim 15$ eV). At higher electron energies the momentum transfer in electron-electron collisions is negligible in comparison to energy loss in inelastic collisions.  The the ratio of electron-electron collisions to inelastic collisions with neutrals depends on the ionization degree and temperature of the electron population \cite{BoltzmannSolver}.

The plasma heating method affects the volume production efficiency of negative hydrogen ions. If electrons are heated rapidly to high energies, their energy is dissipated in reaction favoring the volume production i.e. ionization and excitation to $B^1\Sigma^+_u$ and $C^1\Pi_u$ states. The resulting electron energy distribution consists of a (dominating) cold population and an energetic tail. This is the case in filament driven arc discharges, which are efficient in producing negative hydrogen ions. If electrons are not heated to high energies quickly, the electron energy dissipates in electron-electron collisions and unfavorable inelastic collisions, e.g. excitation to triplet states, during the heating process. Under such conditions the heating power does not only dissipate in unfavorable processes for the volume production but the average energy of the electrons in the filter field region might also increase, which is harmful for negative ions. From this point of view, ECR heating should be more suitable for the volume production than off-resonance heating.

\section{Experimental setup}

\subsection{Microwave discharge}
\begin{figure}[ht!]
 \centering
 \includegraphics[width=0.85\textwidth]{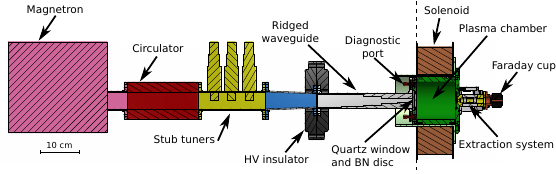}% % Important NOTE: Please make certain your figures do not %include local directory paths. ex. "c:\file\sub\fig1.eps"
 \caption{\label{fig:microwave_source} The microwave ion source. The microwave coupling system (on the left), which was connected to the LIISA ion source and the RADIS ion source, is separated with a dashed line.}%
 \end{figure} 
The LIISA ion source and the RADIS ion source were operated in microwave driven mode by using the microwave coupling system of a microwave ion source. The waveguide-based microwave coupling system connected to a discharge chamber of the microwave ion source is presented in Fig \ref{fig:microwave_source}. The ion source consists of an extraction system for positive ions, a cylindrical aluminium plasma chamber which is surrounded by a single solenoid coil, and the microwave coupling system. The vacuum of plasma chamber is sealed from the waveguide (under atmospheric pressure) by a quartz window embedded into the back plate of plasma chamber. The ion source is typically operated with 300-1200 W of microwave power in 0.05 Pa to 10 Pa pressure range.

\begin{figure}[ht!]
 \centering
 \includegraphics[width=0.9\textwidth]{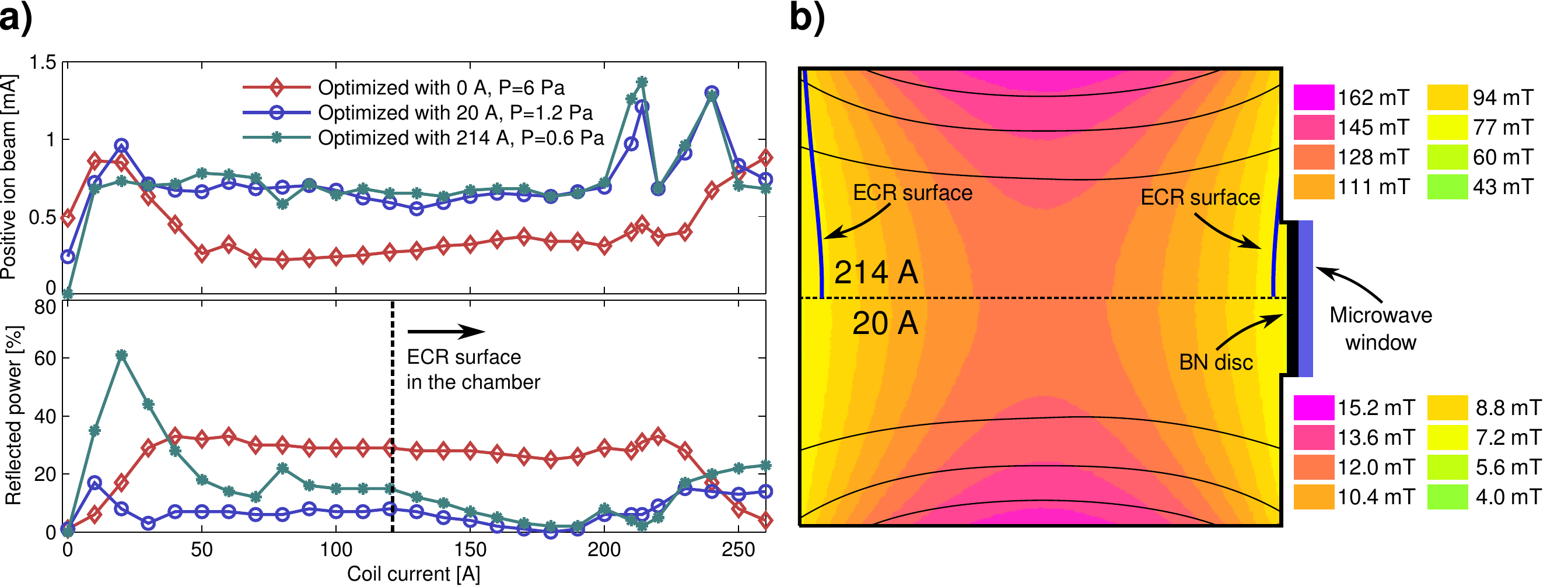}% % Important NOTE: Please make certain your figures do not %include local directory paths. ex. "c:\file\sub\fig1.eps"
 \caption{\label{fig:microwave_performance} \textbf{a)} Total positive (H$^+$, H$_2^+$ and H$_3^+$) ion beam current as a function of the solenoid coil current. The ion beam was optimized in three different cases: without magnetic field, with 20 A coil current and with optimum resonance magnetic field (214 A). The incident microwave power was 1.2 kW and the beam was extracted from a 1.5 mm diameter hole with an extraction voltage of 12.5 kV. The maximum current density is comparable to performances of the similar type of microwave ion sources \cite{BNdisc,microwave_performance2,microwave_performance3}. \textbf{b)} A cross sectional view of the magnetic field structure with 20 A and 214 A coil currents. The magnetic fields have been calculated with FEMM 4.2 \cite{femm}.}%
 \end{figure}

The simple magnetic field structure of the microwave ion source allows a straightforward study of the electron heating process. Three plasma heating regimes (1. no magnetic field, 2. off-resonance and 3. electron cyclotron resonance) are demonstrated in Fig. \ref{fig:microwave_performance} (a) presenting the ion source performance in terms of positive ion beam production as a function of the magnetic field strength. The source parameters (pressure and positions of stub tuners) were optimized in three different cases: without magnetic field, with 20 A coil current and with optimum resonance conditions (214 A). The best performance is achieved when the ECR surface is located close to the microwave window, although the performance decreases only 50 \%, at lower magnetic fields. It is also possible to operate the ion source without magnetic field. However, the ignition of the discharge at low pressures (<5 Pa) requires a magnetic field corresponding to a coil current of 120 A or more. The required threshold magnetic field for ignition is lower at higher pressures. The magnetic field dependence of the source performance is sensitive to the settings of the impedance tuners.

A boron nitride (BN) disc is typically used to protect the quartz window from backstreaming electrons from the extraction system. It has been observed, that the boron nitride disc improves the proton fraction of the extracted beam \cite{BNdisc}. There are several hypotheses for the effect, such as secondary electron emission from BN and plasma chemical processes \cite{BNdisc}. In the source discussed here, the boron nitride disc (thickness 4 mm) covers the quartz window.

\subsection{Filament driven arc discharge LIISA}
\begin{figure}[ht]
 \centering
 \includegraphics[width=0.75\textwidth]{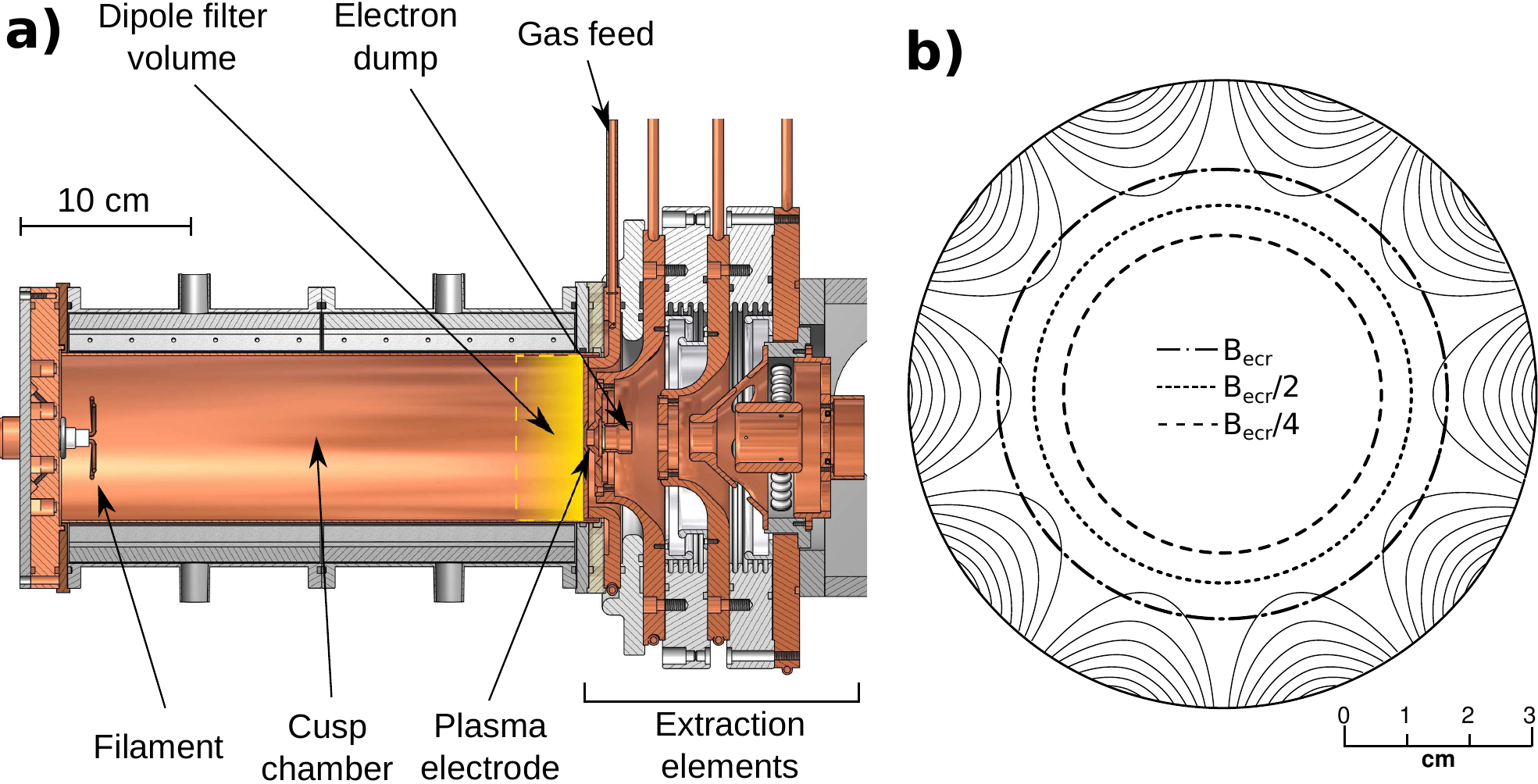}% % Important NOTE: Please make certain your figures do not %include local directory paths. ex. "c:\file\sub\fig1.eps"
 \caption{\label{fig:liisa} \textbf{a)} The LIISA ion source \textbf{b)} Constant magnetic field surfaces of B$_{ecr}$, B$_{ecr}$/2 and B$_{ecr}$/4. The magnetic fields have been calculated with FEMM 4.2 \cite{femm}. }%
 \end{figure}
 
The filament driven arc discharge negative ion source LIISA is presented in Fig. \ref{fig:liisa} (a). It consists of a decapole (10-pole) multicusp plasma chamber, back plate housing two tantalum filaments, biased plasma electrode and extraction system. The diameter and the length of the plasma chamber are 9 cm and 31 cm, respectively. The plasma chamber is made of copper, but it is coated with tantalum due to the evaporation of the filament. More detailed description of the ion source can be found from Refs. \cite{LIISA,komppula_NIBS_2012}.

Typical operational parameters of LIISA are 500-1000 W of arc discharge power and 0.2-0.8 Pa pressure. The ion source can provide an H$^-$-beam of 1.4 mA extracted through 11 mm aperture with 1 kW discharge power. The maximum beam (> 1mA, 1 kW) can be achieved with the co-extracted electron / H$^{-}$ ratio varying from 16 (0.2 Pa) to less than 2 (>0.5 Pa).

The multicusp magnetic field configuration of the LIISA plasma chamber is presented in Fig. \ref{fig:liisa} b). The strength of the magnetic field on the chamber wall is 350 mT at the poles and 290 mT between the poles. The multicusp magnetic field provides cylindrical surfaces of constant magnetic field. The distance of the 2.45 GHz resonance field (87.5 mT) from the chamber wall is 1.4 cm. For half resonance field (43.8 mT) the corresponding distance is 2.0 cm.

The LIISA ion source was converted to microwave driven configuration by replacing the back plate housing the filaments with the back plate of the microwave source.

\subsection{RF-driven discharge RADIS}
\begin{figure}[ht]
 \centering
 \includegraphics[width=0.8\textwidth]{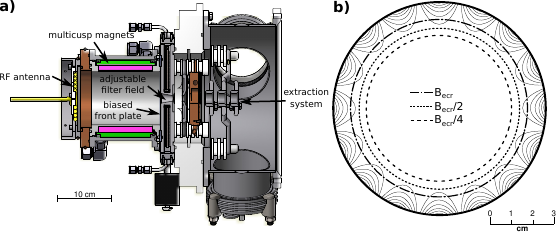}% % Important NOTE: Please make certain your figures do not %include local directory paths. ex. "c:\file\sub\fig1.eps"
 \caption{\label{fig:radis} \textbf{a)} The RADIS ion source \textbf{b)} Constant magnetic field surfaces of B$_{ecr}$, B$_{ecr}$/2 and B$_{ecr}$/4. The magnetic fields have been calculated with FEMM 4.2 \cite{femm}.}%
 \end{figure}

The RF-driven RADIS ion source is presented in Fig. \ref{fig:radis}. It consists of a 16-pole multicusp plasma chamber, AlN back plate with an external spiralled planar antenna and extraction system. The antenna is connected to the 13.56 MHz RF-generator via capacitive matching network. The plasma chamber is made of aluminium. The diameter and length of the plasma chamber are 98 and 137 mm, respectively. More detailed description of the RADIS ion source can be found from Refs. \cite{NIBS2012_Kalvas,NIBS2014_Kalvas}.

Typical operational parameters of RADIS are 500-2000 W of RF-power at \SIrange[per-mode=fraction]{0.5}{1}{\pascal} pressure. Typically the RADIS ion source can produce 0.3 mA of H$^-$ through 7 mm aperture with 1 kW of RF power. The ratio of co-extracted electrons to H$^-$ is typically 10--30. The plasma ignition requires pressure of 4.5 Pa.

The multicusp magnetic field configuration of the RADIS plasma chamber is presented in Fig. \ref{fig:radis} b). The strength of the magnetic field on the chamber wall is 340 mT at the poles and 290 mT between the poles. The distance of the 2.45 GHz resonance field (87.5 mT) from the chamber wall is 0.8 cm. For half resonance field (43.8 mT) the corresponding distance is 1.2 cm.

\begin{figure}[h]
 \centering
 \includegraphics[width=0.90\textwidth]{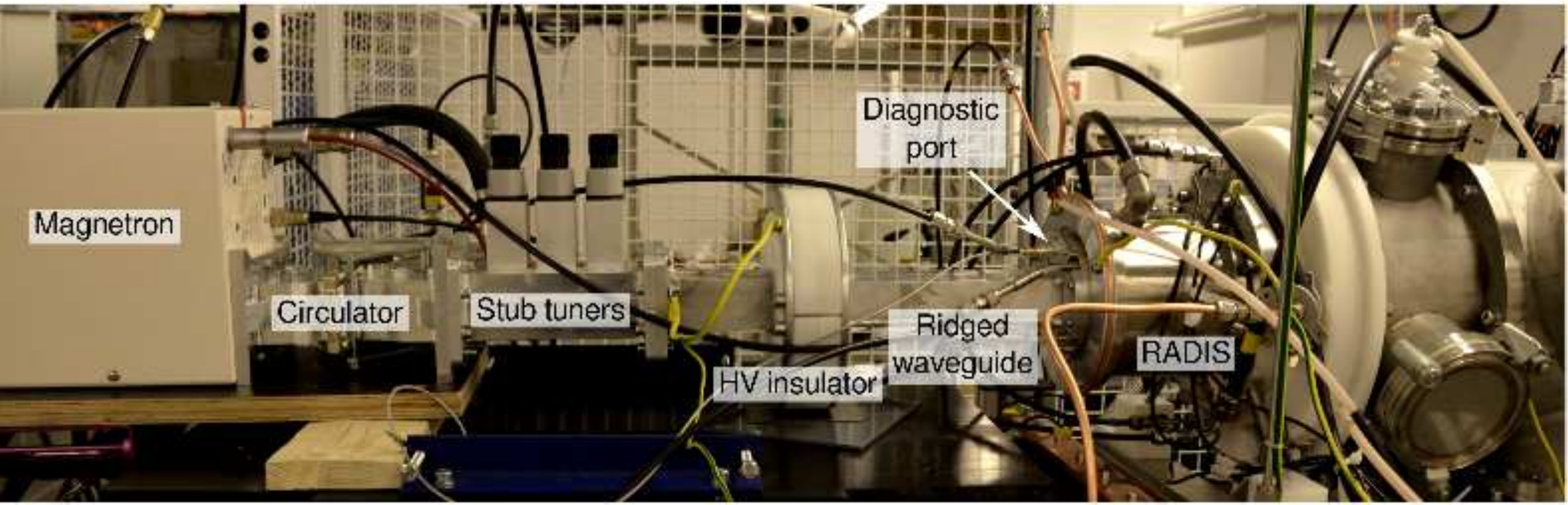}% % Important NOTE: Please make certain your figures do not %include local directory paths. ex. "c:\file\sub\fig1.eps"
 \caption{\label{fig:radis_photo} Microwave coupling system of the microwave discharge ion source (Fig. \ref{fig:microwave_source}) connected to the plasma chamber and the extraction system of the RADIS ion source.}%
 \end{figure}

The RADIS ion source was converted to microwave driven configuration by replacing the AlN back plate of the plasma chamber with the back plate of the microwave ion source plasma chamber with an adapter (Fig. \ref{fig:radis_photo}). 

\section{Results}
Since the described experiments were performed for the first time, several challenges limiting the parameter space of source tuning were encountered. Therefore, the comparison of different heating methods is performed by using heating power of one kilowatt. The highest beam currents and optimized source tuning parameters for different heating methods and plasma chambers are presented in Table \ref{tab:results}. 

The results indicate that the heating efficiency achieved with microwaves is at least 30\% of the efficiency achieved with the heating methods dedicated to each discharge chamber. The tests imply that the optimum pressure range and amount of the co-extracted electrons does not significantly depend on the heating method. The most significant difference was observed in the optimum voltage of the biased plasma electrodes. In the case of the LIISA, the optimum voltage was approximately 10 V when microwave heating was used and 3 V in the filament-driven mode. In the case of RADIS, the optimum H$^-$ current was obtained when the bias voltage was from 0 V to reversed.

\begin{table}
\begin{tabular}{lrrrrl}  
\hline
  & \tablehead{1}{c}{b}{RADIS\\RF}
  & \tablehead{1}{c}{b}{RADIS\\2.45 GHz }
  & \tablehead{1}{c}{b}{LIISA\\filament}
  & \tablehead{1}{c}{b}{LIISA\\2.45 GHz}
  & \tablehead{1}{c}{b}{}   \\
\hline
H$^-$ beam current (CW)& \SI[per-mode=fraction]{300}{} & \SI[per-mode=fraction]{170}{} & \SI[per-mode=fraction]{1500}{} & \SI[per-mode=fraction]{500}{}& \SI[per-mode=fraction]{}{\micro\ampere}\\
H$^-$ current density& \SI[per-mode=fraction]{0.78}{} & \SI[per-mode=fraction]{0.44}{} & \SI[per-mode=fraction]{1.6}{} & \SI[per-mode=fraction]{0.53}{}& \SI[per-mode=fraction]{}{\milli\ampere\per\centi\meter^2}\\
e/H$^-$ & 10--30 & 25--35 & 2--4 & 5--15\\
Pressure  & 0.5--1 & \SI[per-mode=fraction]{1}{} & \SI[per-mode=fraction]{0.45}{} & \SI[per-mode=fraction]{0.65}{}& \SI[per-mode=fraction]{}{\pascal}\\
Plasma electrode & 7--12 & 0 & 1--4 & 2--15 & \SI[per-mode=fraction]{}{\volt}\\
\hline
%&&&&\\
\textbf{Ignition thresholds } &  &  &  & \\
Pressure  & $\approx$\SI[per-mode=fraction]{4.5}{} & $\approx$\SI[per-mode=fraction]{4.5}{} & - & \SI[per-mode=fraction]{0.75}{}& \SI[per-mode=fraction]{}{\pascal}\\
Power    & $\approx$\SI[per-mode=fraction]{1.0}{}  & $\approx$\SI[per-mode=fraction]{1.5}{} & - & $\approx$\SI[per-mode=fraction]{1.5}{}& \SI[per-mode=fraction]{}{\kilo\watt}\\
\hline
\end{tabular}
\caption{\label{tab:results} Measured H$^-$ beam currents and tuning parameters for different ion sources and heating methods. A heating power of 1 kW has been used in each case. The reported beam currents are achieved with pure hydrogen. The boron nitride disc was used in front of the quartz microwave window in the case of RADIS, but not in the case of LIISA.}
\end{table}

Two different 'plasma modes' were observed with both plasma chambers when microwave heating was applied. The color difference of the 'plasma modes' was caused by varying ratio of atomic and molecular light emission. Atomic emission, dominated by the Balmer-alpha line in red part of the visible light spectrum, was significantly more intensive in the 'red plasma mode' than the wide band of light emission of molecular hydrogen. The best performance was achieved with 'red plasma mode'. The ignition of this mode required at least 1.5 kW of microwave power and higher neutral pressure. Once the plasma was ignited, it was possible to operate both sources with lower pressure and power (> 900 W) and sustain the 'red mode'. The other plasma mode was a 'blue mode', which ignited with low microwave power and low pressure. The performance in this mode was not as good as in the 'red mode', when 1 kW of heating power was used. However, \SI[per-mode=fraction]{240}{\micro\ampere} of H$^-$ was achieved with 130 W of heating power (1.01 kW forward and approximately 880 W reflected) with the plasma chamber of LIISA. Unfortunately it was impossible to obtain such power efficiency with lower reflected power. 

The maximum microwave power was limited by overheating of the back plate. The temperature of the ridged waveguide was $80-100\,^{\circ}\mathrm{C}$ at 1 kW microwave power. At higher power the H$^-$ current vanished, which is supposed to be due to outgassing or evaporation of impurities from the back plate. The given explanation is supported by the fact that unknown peaks appeared in the visible light spectrum when the H$^-$ production deteriorated at high power levels. Repeated overheating coated the plasma chamber with a layer of impurities. Signs of melting were found from the quartz window after some hours of operation at high power level. This indicates that the temperature of the window was some hundreds of $^{\circ}\mathrm{C}$. The impurity layer decreased the source performance by approximately 50\%. The performance of the LIISA chamber was restored by evaporating tantalum from the filament to the plasma chamber surfaces. Restoring the RADIS performance required mechanical polishing of the surfaces.

The effect of mixing noble gas (argon) with hydrogen was tested with both plasma chambers. Up to 50\% increase of negative ion beam current was achieved in both cases at lower power. However, injection of the noble gas made the overheating problems more severe. The effect of krypton was similar to argon in the case of LIISA but was not tested with RADIS.

\section{Discussion}
The main purpose of the described tests was to demonstrate the potential of microwave heating in negative ion source technology by using plasma chambers and extraction systems of conventional multicusp ion sources. The result, namely the 30\% performance in comparison to nominal plasma heating method for each chamber, is very promising especially because the systems are far from being optimized for microwaves. Furthermore, a similar test was performed earlier on LIISA with RF-heating and the performance was only \SI[per-mode=fraction]{240}{\micro\ampere} \cite{NIBS2012_Kalvas}, half of the performance achieved with microwave heating.

The most critical part of the experimental setups used in this study is the back plate of the plasma chamber. Since the back plate has no magnets embedded into the structure, it is the weakest point from the confinement point-of-view. Thus, the cooling system of the back plate should be designed to be capable of handling all the plasma heating power. The quartz window is clearly not the optimum technical solution. Alumina (Al$_2$O$_3$) and aluminum nitride (AlN) are promising candidates as they are proven to withstand tens of kilowatts of power in pulsed operation \cite{AlO3Chamber,AlNChamber} and several kilowatts in continuous operation \cite{NIBS2012_Kalvas,NIBS2014_Kalvas}.

The existence of two 'plasma modes' implies that the power absorption mechanism is somewhat different in each case. Sustaining the 'red mode' requires high power and higher pressure, which are also the required conditions for running the microwave discharge without magnetic field (Fig. \ref{fig:microwave_performance} (a)). This could imply that under such conditions the microwave heating occurs mainly via an off-resonance process. On the contrary, the 'blue mode' could be a result of ECR heating. This is because, it can be achieved with a wide range of parameters (pressure and power), but requires the presence of ECR-surface. The weaker performance in the 'blue mode' could be due to stronger binding of electrons into the magnetic field lines. In the cusp configurations the resonance surfaces of ECR heating are close the wall of the plasma chamber (Figs. \ref{fig:liisa} (b) and \ref{fig:radis} (b)), which means that most of the heated electrons are bound to the field lines near the wall. Therefore, a long diffusion length in transverse direction with respect to the magnetic field is required for the plasma to reach the extraction region. Furthermore, only the lowest cavity modes of electric fields ($\lambda_{2.45 GHz}=0.12$ m) can be excited in the plasma chamber (d=\SI[per-mode=fraction]{95}{\milli\meter}, L=\SIrange[per-mode=fraction]{100}{300}{\milli\meter}) \cite{MicrowaveCoupling} and therefore the electric field is weak on the resonance surface.

An optimized plasma chamber for negative ion production could offer not only adequate conditions for ECR heating but also improved plasma diffusion towards the extraction region. Reducing the number of multicusp poles would allow utilizing a larger volume of efficient ECR heating closer to the axis of the plasma chamber \cite{ECRIN_multicusp_electric_filter,MulticuspSimulation} and better integration of the magnetic filter field with the cusp-field. The drawback of such approach might be weaker plasma confinement in the radial direction.

To overcome the limitations of increasing plasma density, i.e. reaching the cut-off plasma density, it is possible to use higher microwave frequencies. Extraction of positive ion beams (up to 600 mA/cm$^2$ of H$^+$) from hydrogen discharge with microwave frequencies up to 37.5 GHz and heating power of 100 kW has been demonstrated \cite{protoni_morko}. This shows that technical limitations of plasma heating are far from 2.45 GHz microwave frequency. However, the referred results cannot be directly compared to the production efficiency of negative hydrogen ions for which the plasma density is not the only relevant parameter but other processes such as plasma chemistry and plasma diffusion play an important role.

It can be argued that either a hexapole or modified quadrupole magnetic field together with a plasma chamber with a diameter corresponding approximately to the wavelength of higher frequency microwaves (5.8 GHz to 14 GHz) could be an optimum solution. The magnetic filter field could be easily optimized and integrated to the topology of the multicusp magnetic field. Furthermore, high plasma densities could be achieved with reasonably small power which is allowed by the high cut off density of microwaves and small plasma volume. The most severe drawback of such configuration might be the small size of the plasma chamber that could lead to decreased density of molecules on high vibrational levels due to pronounced molecule-wall interactions. 

The reported results imply that, if the source performance in terms of maximum current is not critical, traditional filament driven arc discharges can be straightforwardly converted into almost maintenance free microwave driven ion sources by merely replacing the back plate of the source. Higher beam currents could be achieved by improving the design of the back plate. For example, enhancing the cooling would allow operating the ion source at higher power levels. In addition, implementation of a weak solenoid field (Fig. \ref{fig:microwave_performance} (a)) could improve the plasma heating, confinement and density.

Our future plan is to design a new biased front plate to the RADIS ion source \cite{NIBS2014_Kalvas}. This allows comparison of different heating methods (filament arc, RF, microwaves) on a single plasma chamber equipped with conventional magnetic filter field. Such study would provide new information on the benefits and drawbacks of different heating methods in relation to the volume production of negative ions.

%%%%%%%%%%%%%%%%%%%%%%%%%%%%%%%%%%%%%%%%%%%%%%%%
%% BACKMATTER
%%%%%%%%%%%%%%%%%%%%%%%%%%%%%%%%%%%%%%%%%%%%%%%%

\begin{theacknowledgments}
  This work has been supported by the EU 7th framework programme 'Integrating
Activities -- Transnational Access', project number: 262010 (ENSAR) and by the
Academy of Finland under the Finnish Centre of Excellence Programme 2012--2017
(Nuclear and Accelerator Based Physics Research at JYFL)
\end{theacknowledgments}



\bibliographystyle{aipproc}   % if natbib is available

\bibliography{NIBS2014_komppula}


\end{document}